\documentclass[
 reprint,
 superscriptaddress,
 amsmath,
 amssymb,
 aps,
 prl
]{revtex4-2}

\usepackage{floatrow}
\usepackage{graphicx}
\usepackage[utf8]{inputenc}
\usepackage{color}
\usepackage{hyperref}

\renewcommand{\t}{\boldsymbol} 
\renewcommand{\tt}{\mathbb} 

\begin{document}

\title{Efficient topology optimization using compatibility projection in micromechanical homogenization}
\author{Indre Jödicke}
\affiliation{Department of Microsystems Engineering, University of Freiburg, Georges-K\"ohler-Allee 103, 79110 Freiburg, Germany}
\affiliation{Cluster of Excellence livMatS, Freiburg Center for Interactive Materials and Bioinspired Technologies, University of Freiburg, Georges-K\"ohler-Allee 105, 79110 Freiburg, Germany}

\author{Richard J. Leute}
\affiliation{Department of Microsystems Engineering, University of Freiburg, Georges-K\"ohler-Allee 103, 79110 Freiburg, Germany}

\author{Till Junge}
\affiliation{Laboratory for Multiscale Mechanics Modeling, Institute of Mechanical Engineering, School of Engineering, \'Ecole Polytechnique F\'ed\'erale de Lausanne, 1015 Lausanne, Switzerland}

\author{Lars Pastewka}
\affiliation{Department of Microsystems Engineering, University of Freiburg, Georges-K\"ohler-Allee 103, 79110 Freiburg, Germany}
\affiliation{Cluster of Excellence livMatS, Freiburg Center for Interactive Materials and Bioinspired Technologies, University of Freiburg, Georges-K\"ohler-Allee 105, 79110 Freiburg, Germany}

\begin{abstract}
    The adjoint method allows efficient calculation of the gradient with respect to the design variables of a topology optimization problem. This method is almost exclusively used in combination with traditional Finite-Element-Analysis, whereas Fourier-based solvers have recently shown large efficiency gains for homogenization problems. In this paper, we derive the discrete adjoint method for Fourier-based solvers that employ compatibility projection. We demonstrate the method on the optimization of composite materials and auxetic metamaterials, where void regions are modelled with zero stiffness.

\end{abstract}

\maketitle

\section{Introduction}

In 1988 Bends{\o}e and Kikuchi published their seminal work on topology optimization~\cite{Bendsoe1988} and seven years later, Sigmund applied topology optimization to design materials with prescribed elastic properties~\cite{Sigmund1994, Sigmund1995}. Since then tailoring material properties with topology optimization has evolved into a large research area (see Ref.~\cite{Osanov2016} for a review). One ongoing research topic is to reduce the high computational cost of topology optimization problems (e.g.~Refs.~\cite{Sigmund2013, Peetz2021}). A large amount of this cost comes from the solution of the physical equilibrium, usually carried out with Finite-Element-Analysis. As an alternative, Fourier-based solvers show promise in becoming an efficient tool for the solution of the static mechanical homogenization problems~\cite{Roters2012}. Fourier-based solvers have not yet been employed in the context of topology optimization, and we demonstrate their efficiency and versatility in the present article.

Fourier-based solvers for mechanical equilibrium trace back to the works of Moulinec and Suquet in the nineties~\cite{Moulinec1994, Moulinec1998}. Since then, Fourier-based solvers have greatly evolved (see~Ref.~\cite{Schneider2021} for a current review) and have been successfully used for complex homogenization problems, e.g.~Refs.~\cite{deGeus2016, Lucarini2019}. Fourier-based solvers rely on efficient Fast-Fourier-Transformation implementations (like FFTW~\cite{Frigo2005}) for computational efficiency. A variant of the method using \emph{compatibility projection} was developed by Vond\-{r}ejc, Zeman, de Geus et al. during the last decade~\cite{Vondrejc2014,Zeman2017,deGeus2017}. These solvers work with some strain measure (e.g. the deformation gradient) rather than the displacements, as the unknown variable. The compatibility of this strain measure is enforced by projection, so that no additional compatibility equation must be solved. Here we will use the strain-based finite-element projection recently proposed by us~\cite{Leute2022} to avoid ringing problems, a phenomenon of artificial oscillations around discontinuities \cite{Gottlieb1997} persisting in many Fourier-based solvers (e.g. \cite{Schneider2016, Ma2021}) and to  enable modeling of internal free surfaces. For the topology optimization context, this means that void can be modelled with zero stiffness and does not need to be approximated by a very weak material.

In this article we show how to use a Fourier-based solver with compatibility projection in gradient-based topology optimization. There are different approaches to formulate the topology optimization problem as a well-posed, differentiable optimization problem (e.g. Refs.~\cite{Bendsoe2003, Sigmund2013}). We will use the phase-field method proposed by Bourdin et al. in 2003~\cite{Bourdin2003} and since then successfully used in different optimization problems~\cite{Wang2004, Burger2006, Wallin2012, Dondl2019}. To calculate the sensitivity, i.e. the gradient with respect to the design parameters, we will work with the discrete adjoint method. The adjoint method is the most efficient way to perform a sensitivity analysis of problems with many design variables (e.g.~Ref.~\cite{Tortorelli1994}). It has therefore become a standard for Finite-Element-Method based topology optimization problems (e.g.~Refs.~\cite{Bendsoe2003, Osanov2016}). Nevertheless, to the best of our knowledge, the adjoint method has not yet been formulated for a Fourier-based solver.

As exemplary topology optimization problem we have chosen two relatively simple, small-strain 2D problems: First we optimize a composite material for a specific shear modulus; second we optimize an auxetic metamaterial with negative Poisson's ratio.

\section{Equilibrium solver with compatibility projection}
\label{sec:equilibrium}
In this section, we will briefly recapitulate the theory of a Fourier-based solver with compatibility projection.

We are working in a periodic unit cell of (undeformed) volume $\Omega_0$ that contains our representative volume element. The aim is to solve the static mechanical equilibrium given by
\begin{equation}
\label{eq:equilibrium}
    \nabla \cdot \t{P}^T(\t{F}(\vec{X})) = \vec{0}
\end{equation}
where $\nabla$ is the nabla operator with respect to the undeformed positions $\vec{X}$, the operator $\cdot$ is the dot product $[\vec{a} \cdot \t{B}]_j = a_i B_{ij}$ where here and in what follows we assume implicit summation over repeated indices (Einstein summation convention). $\t{P}(\vec{X})$ is the first Piola-Kirchhoff stress and the superscript $T$ designates the tensor transpose. Here and in the following bold symbols represent second order tensors and arrows represent a vector. $\t{P}$ depends on the local deformation gradient $\t{F}(\vec{X})$ through an appropriate constitutive expression. $\t{F}(\vec{X})$ is defined by $F_{ij}=\partial x_i/\partial X_j$ where $x_i$ is the deformed position.

We now formulate the weak form of the equilibrium equation following the procedure outlined in Ref.~\cite{deGeus2017}. Given a periodic test function $\vec{v}$, we rewrite Eq.~\eqref{eq:equilibrium} in the weak form,
\begin{equation}
    \label{eq:equilibrium_weak_form}
    \begin{split}
    \int_{\Omega_0} d \Omega_0 \; \vec{v} \cdot \left( \nabla \cdot \t{P}^T\left(\t{F}\right)\right) &= - \int_{\Omega_0} d \Omega_0 \; \left( \nabla \otimes \vec{v}\right) : \t{P}^T\left( \t{F}\right) \\
    &= 0,        
    \end{split}
\end{equation}
where we have applied the divergence theorem and used the periodic boundary conditions to eliminate the surface term. The operator $:$ designates the double contraction, $\t{A}:\t{B}=A_{ij}B_{ji}$. The operator $\otimes$ denotes the outer product, i.e. $[\vec{a} \otimes \vec{b}]_{ij} = a_i b_j$. The tensor $\nabla \otimes \vec{v}$ can be interpreted as a virtual deformation gradient $\delta \t{F}$, i.e. a periodic and \emph{compatible} test function. By \emph{compatible} we mean, that it is given by the gradient of a potential function.

Compatibility can be enforced by a convolution with the self-adjoint compatibility projection operator $\mathbb{G}$ (see e.g. Refs.~\cite{Vondrejc2014, Leute2022}), so that we can write:
\begin{equation}
\label{eq:equilibrium_and_compatibility}
\begin{split}
    \int_{\Omega_0} d \Omega_0 \; \delta \t{F} : \t{P}^T\left( \t{F}\right) =&
    \int_{\Omega_0} d \Omega_0 \; \left( \mathbb{G} \star \delta \t{\Tilde{F}}\right) : \t{P}^T\left( \t{F}\right) \\
    =& \int_{\Omega_0} d \Omega_0 \; \delta \t{\Tilde{F}} : \left( \mathbb{G} \star \boldsymbol{P}\left( \boldsymbol{F}\right)\right)^T
    \\
    &= 0,
\end{split}
\end{equation}
where $\star$ is the application of the projection operator to the right hand-side and $\delta \t{\Tilde{F}}$ a periodic test function, which no longer has to be compatible. The operation $\star$ is a convolution in real-space but becomes a multiplication in Fourier space. The numerical methods behind this compatibility projection therefore employ fast Fourier transforms to accelerate the computation of $\mathbb{G}\star$ (see Refs.~\cite{deGeus2017, Zeman2017, Leute2022, Ladecky2022}).

Eq.~\eqref{eq:equilibrium_and_compatibility} is discretized on a regular grid using a Galerkin scheme, i.e. the test functions $\delta \t{\Tilde{F}}$ and the unknown $\t{F}$ are expressed with the same set of basis functions. This leads to the discretized equilibrium equation:
\begin{equation}
\label{eq:equilibrium_discrete}
    \tt{G}_{\alpha \beta}:\t{P}_{\beta}\left( \t{F}\right) = \t{0}
\end{equation}
where small greek letters refer to the element of the discretized domain. Note that in the discretized form, $\mathbb{G}$ is a fourth order tensor and the double contraction yields $[\tt{G}:\t{P}]_{ij}=G_{ijkl} P_{lk}$, where we have omitted the greek element indices for brevity.

The projection operator $\tt{G}$ projects every tensor field onto its compatible part. Suitable definitions employing Fourier-derivatives for $\nabla$ are given for finite-strain in Ref.~\cite{deGeus2017} and for small-strain in Ref.~\cite{Zeman2017}. In Ref.~\cite{Leute2022}, the formulations for both cases are extended for arbitrary discretization of the gradient operator. In all cases $\tt{G}$ is block-diagonal in Fourier-space, so that Eq.~\eqref{eq:equilibrium_discrete} is always evaluated in Fourier-space for the sake of numerical efficiency. Please note that whereas the specific form of $\tt{G}$ varies, the form of Eq.~\eqref{eq:equilibrium_discrete} remains the same and the derivation in section \ref{sec:adjoint_method} is independent of the specific form of $\mathbb{G}$ that is used. The only properties of $\mathbb{G}$ that are required are its self-adjointness and that it projects every tensor field unto its compatible part.

Equation~\eqref{eq:equilibrium_discrete} is usually nonlinear. It must therefore be solved iteratively, e.g. by Newton iteration (see Ref.~\cite{deGeus2017}):
\begin{equation}
\label{eq:equilibrium_newton_iteration}
\begin{aligned}
&\t{F}_{(i+1),\alpha} = \t{F}_{(i),\alpha} + \Delta \t{F}_{\alpha} \\
&\text{with} \quad \tt{G}_{\alpha\beta}:\tt{K}^{LT}_{(i),\beta\gamma}:\Delta \t{F}^T_{\gamma} = -\tt{G}_{\alpha\beta}: \t{P}_{(i),\beta}
\end{aligned}
\end{equation}
Here $\tt{K}^{LT}$ is the assembly of the left-transposed, i.e. $[\tt{K}^{LT}]_{ijkl} = K_{jikl}$, of the local tangent stiffness tensor $\tt{K}$, defined by the linearization of the stresses around $\t{F}_{(i)}$, $d \t{P}^T = \tt{K} : d \t{F}^T$, and $i$ denotes the step of the Newton iteration. The tangent stiffness tensor $\tt{K}$ is block-diagonal, so that the double dot product in Eq.~\eqref{eq:equilibrium_newton_iteration} can be efficiently implemented. As shown in Refs.~\cite{Vondrejc2014, Mishra2016, Zeman2017}, the increments $\Delta \t{F}$ are themselves compatible, which ensures the compatibility of the deformation gradient obtained by this iterative procedure. It is therefore not necessary to solve an equation for satisfying compatibility of the deformation gradient.

\section{Discrete adjoint method}
\label{sec:adjoint_method}
In this section, we will formulate the discrete adjoint method for topology optimization problems using a Fourier-based solver with compatibility projection.

The aim of the discrete adjoint method is to calculate the sensitivity $S$ (the gradient with respect to the design variable) of the optimization problem in the discretized form:
\begin{equation}
\label{eq:optimization_problem}
\begin{aligned}
& \underset{\rho}{\text{minimize}}
&&f\left(\rho, \t{F}_{1}, \t{F}_{2}, ..., \t{F}_{M} \right) \\
& \text{subject to}
&& \tt{G}_{\alpha \beta}:\t{P}_{\Gamma,\beta} = \t{0} \qquad \forall \, \Gamma \in {1, 2, ..., M}
\end{aligned}
\end{equation}
with $f$ as the aim function and $M$ as the number of equilibrium constraints to which the optimization problem is subjected, typically differing in boundary condition only. Here and in the following, a capital Greek letter index refers to the number of equilibrium constraints and a comma in the index serves to separate different kind of indices for readability. $\rho$ is the design variable, in our case the material density. Note that the problem given in Eq.~\eqref{eq:optimization_problem} is in the discretized form.

\subsection{Derivation of the adjoint method}
For the derivation of the adjoint method, we follow Refs.~\cite{Tortorelli1994, Giles2000}. We introduce a set of second order tensors $\t{\Lambda}_{\Gamma}$ as Lagrangian multipliers to fulfill the constraints. The optimization problem Eq.~\eqref{eq:optimization_problem} is then equivalent to
\begin{equation}
\label{eq:opt_problem_Lagrange}
    \underset{\rho}{\text{minimize}} \; f + \t{\Lambda}_{\Gamma, \alpha}^* : \left( \tt{G}_{\alpha \beta}:\t{P}_{\Gamma,\beta} \right)^T
\end{equation}
as long as the equilibrium constraints are satisfied separately. The superscript $^*$ designates the complex conjugate of a variable. Since the design parameters are independent of each other, the definition of the sensitivity at element $\gamma$ leads to:
\begin{equation}
\label{eq:def_sensitivity}
\begin{aligned}
S_{\gamma} = \frac{df}{d \rho_{\gamma}} =& \frac{\partial f}{\partial \rho_{\gamma}} + \t{\Lambda}_{\Gamma, \alpha}^* :  \frac{\partial \left( \tt{G}_{\alpha \beta}:\t{P}_{\Gamma,\beta} \right)^T}{\partial \rho_{\gamma}} \\
&+ \left( \frac{\partial f}{\partial \t{F}_{\Theta, \zeta}} + \t{\Lambda}_{\Gamma, \alpha}^* : \frac{\partial \left( \tt{G}_{\alpha \beta}:\t{P}_{\Gamma,\beta} \right)^T}{\partial \t{F}_{\Theta, \zeta}}\right) : \left( \frac{d \t{F}_{\Theta, \zeta}}{d \rho_{\gamma}} \right)^T
\end{aligned}
\end{equation}
The most complicated terms in this equation are the total derivatives of the strains with respect to the material density. They are defined implicitly by the equilibrium constraints, i.e. they can not usually be calculated analytically. To avoid a costly numerical computation of these terms, we choose $\t{\Lambda}_{\Gamma}$ so that the expression in the parenthesis in front of these terms vanishes.

In other words, $\t{\Lambda}_{\Gamma}$ must fulfill the adjoint equations:
\begin{equation}
\begin{aligned}
    -\frac{\partial f}{\partial \t{F}_{\Theta, \zeta}} &= \t{\Lambda}_{\Gamma, \alpha}^* : \frac{\partial \left( \tt{G}_{\alpha \beta}:\t{P}_{\Gamma,\beta} \right)^T}{\partial \t{F}_{\Theta, \zeta}} \\
    &= \t{\Lambda}_{\Gamma, \alpha}^* : \left( \tt{G}_{\alpha \beta}:\frac{\partial \t{P}_{\Gamma,\beta}}{\partial \t{F}_{\Theta, \zeta}}\right)^{LT}
\end{aligned}
\end{equation}
For $\Theta \neq \Gamma$, the partial derivative of the stress with respect to the strain is 0. In the other cases, $\partial \t{P}_{\Gamma,\beta} / \partial \t{F}_{\Theta, \zeta}$ corresponds to the tangent stiffness tensor $\tt{K}$ which is used in Eq.~\eqref{eq:equilibrium_newton_iteration},
\begin{equation}
\label{eq:adjoint_before_complex_conjugate}
\begin{aligned}
    -\frac{\partial f}{\partial \t{F}_{\Theta, \zeta}} = \t{\Lambda}_{\Theta, \alpha}^* :  \left( \tt{G}_{\alpha \beta}:\tt{K}_{\Theta,\beta \zeta}^{LT}\right)^{LT}
\end{aligned}
\end{equation}
where $[\tt{A}:\tt{B}]_{ijkl} = A_{ijmn} B_{nmkl}$ is the double contraction of two forth order tensors.
We confine ourselves to real-valued aim functions. The complex conjugate of Eq.~\eqref{eq:adjoint_before_complex_conjugate} is then:
\begin{equation}
\begin{aligned}
\label{eq:adjoint_derivation}
    -\frac{\partial f}{\partial \t{F}_{\Theta, \zeta}} =  \tt{K}_{\Theta, \zeta \beta}^{LT}:\left(\tt{G}_{\beta \alpha} : \t{\Lambda}_{\Theta, \alpha}\right)^T
\end{aligned}
\end{equation}
since the projection operator and the tangent stiffness tensor are self-adjoint.

$\tt{G}$ projects $\t{\Lambda}_{\Theta}$ to a compatible solution space. In consequence, we can set the non-compatible part of $\t{\Lambda}_{\Theta}$ to zero without loss of generality, so that $\tt{G} : \t{\Lambda}_{\Theta} = \t{\Lambda}_{\Theta}$. By applying $\tt{G}$ to the left-hand side of Eq.~\eqref{eq:adjoint_derivation} we get the final adjoint equations:
\begin{equation}
\label{eq:adjoint_equation}
    -\tt{G}_{\alpha \zeta} : \frac{\partial f}{\partial \t{F}_{\Theta, \zeta}} = \tt{G}_{\alpha \zeta} :  \tt{K}_{\Theta, \zeta \beta}^{LT} : \t{\Lambda}_{\Theta, \beta}^T
\end{equation}

Once the Lagrangian multipliers are determined by Eqs.~\eqref{eq:adjoint_equation}, the sensitivity can be calculated with the remaining terms of Eq.~\eqref{eq:def_sensitivity}:
\begin{equation}
\label{eq:sensitivity_intermediate}
\begin{aligned}
S_{\gamma} =& \frac{\partial f}{\partial \rho_{\gamma}} + \t{\Lambda}_{\Gamma, \alpha}^* : \left( \tt{G}_{\alpha \beta}:\frac{\partial \t{P}_{\Gamma,\beta} }{\partial \rho_{\gamma}} \right)^T \\
=& \frac{\partial f}{\partial \rho_{\gamma}} + \t{\Lambda}_{\Gamma, \alpha}^* : \left( \tt{G}_{\alpha \gamma}:\frac{\partial \t{P}_{\Gamma,\gamma} }{\partial \rho_{\gamma}} \right)^T
\end{aligned}
\end{equation}
where we have used the fact that the stress of an element $\beta$ does not directly depend on the material of an element $\gamma \neq \beta$. Note that the partial derivatives $\partial \t{P}_{\Gamma,\gamma} / \partial \rho_{\gamma}$ can easily be calculated analytically. With the self-adjointness of the projection operator, Eq.~\eqref{eq:sensitivity_intermediate} can be reformulated as:
\begin{equation}
\label{eq:sensitivity_equation}
    \begin{aligned}
    S_{\gamma} =& \frac{\partial f}{\partial \rho_{\gamma}} + \frac{\partial \t{P}_{\Gamma,\gamma} }{\partial \rho_{\gamma}} : \left(\left( \tt{G}_{\gamma \alpha} : \t{\Lambda}_{\Gamma, \alpha} \right)^* \right)^T \\
    =& \frac{\partial f}{\partial \rho_{\gamma}} + \frac{\partial \t{P}_{\Gamma,\gamma} }{\partial \rho_{\gamma}} : \t{\Lambda}_{\Gamma, \gamma}^T
    \end{aligned}
\end{equation}
In the second equality we have made use of the fact that $\t{\Lambda}_{\Gamma}$ are the solution of the adjoint Eqs.~\eqref{eq:adjoint_equation} and therefore compatible and real-valued. Note that the structur of Eq.~\eqref{eq:sensitivity_equation} is similar to the corresponding sensitivity equations in Finite-Element-Analysis based topology optimization problems, see e.g. \cite{Bendsoe2003}.

\subsection{Efficiency of the adjoint method}
The most computationally costly operation in the sensitivity analysis with the discrete adjoint method is the solution of the adjoint Eqs.~\eqref{eq:adjoint_equation}. The adjoint equation is structurally identical to the Newton iteration of the equilibrium solver, Eq.~\eqref{eq:equilibrium_newton_iteration}. The cost of solving one adjoint equation is therefore identical to one Newton iteration. This means the cost of the sensitivity analysis is typically smaller than the cost of solving a (nonlinear) equilibrium problem. An additional advantage of this form of the adjoint equations is that we can use the solver implemented for the Newton-iterations to compute the Lagrange multipliers.

\section{Validation and application}
In this section, we test our adjoint formulation on a 2D small-strain example. We have implemented the method in the open-source Fourier-accelerated micromechanics solver \textsc{\textmu Spectre}~\cite{muspectre}.

\subsection{The test cases}
\label{sec:test_problem}

As test case, we optimize a (periodic) 2D unit cell for a target effective shear modulus in a small-strain situation. As aim function we choose the least square difference between the average stress in the unit cell and the target stress $\t{\sigma}_\text{target}$, which is the Cauchy-stress in the hypothetical homogeneous unit cell with the target material. To ensure isotropy, we consider this error for three linearly independent load cases,
\begin{equation}
    \t{\overline{\epsilon}}_0 = \begin{pmatrix}
    \Delta \epsilon & 0 \\
    0 & 0
    \end{pmatrix} \quad
    \t{\overline{\epsilon}}_1 = \begin{pmatrix}
    0 & 0 \\
    0 & \Delta \epsilon
    \end{pmatrix} \quad
    \t{\overline{\epsilon}}_2 = \begin{pmatrix}
    0 & \Delta \epsilon / 2 \\
    \Delta \epsilon / 2 & 0
    \end{pmatrix},
\end{equation}
where $\t{\overline{\epsilon}}$ is the imposed average small strain and $\Delta \epsilon = 0.01$.

The topology optimization problem needs some form of regularization that picks specific geometries from the variety of geometries that minimize our aim function. We use the phase-field approach of Refs.~\cite{Bourdin2003, Wallin2012}, that selects solutions with minimal interface area. The idea of the phase-field approach is to represent the material distribution by some continuous phase-field function $\rho$ that varies between $0$ and $1$ to represent two materials. The aim function penalizes interfaces through the expression
\begin{equation*}
    \int_0^{L_x} \int_0^{L_y} dy dx \; \left[\eta \vert \nabla \rho \vert^2 + \frac{1}{\eta}\rho^2 \left(1-\rho\right)^2 \right],
\end{equation*}
where $L_x$ ($L_y$) are the length of the unit cell in the first (second) dimension and $\eta$ is a weighting parameter controlling the width of the interface: A smaller $\eta$ penalizes values of $\rho$ between $0$ and $1$, while allowing steeper gradients of $\rho$, so that the width of the diffuse interface decreases as $\eta$ decreases. It can be shown that in the limit $\eta\to 0$, this functional becomes proportional to the total area (or in two-dimensions, length) of the interface~\cite{Modica1977, Modica1987}. In our case, $\eta = 1/40$ of the length of the unit cell in x-direction results in a good interface width.

In the topology optimization context, the double-well potential serves to penalize intermediate terms of $\rho$, i.e. the solution of the continuous optimization problem will converge towards the solution of the discrete optimization problem. The penalization of the gradient avoids the mesh-dependency of the typical topology optimization problem (e.g. \cite{Bendsoe2003} chapter 1.3). Note that, contrary to the 'Solid isotropic material with penalization'-approach often used in topology optimization (e.g. \cite{Bendsoe2003}), the phase-field approach does not require an active volume constraint to converge to discrete solutions. To keep the test cases as simple as possible, we have therefore refrained from adding a volume constraint.

The complete aim function is now:
\begin{equation}
\begin{aligned}
    f(\rho) =& \left(\frac{1}{L_x L_y} \int_0^{L_x} \int_0^{L_y} dy dx \;  \t{\sigma}_{\Gamma}(x,y) - \overline{\t{\sigma}}_{\text{target},\Gamma}\right)^2\\
    &+ w \int_0^{L_x} \int_0^{L_y} dy dx \; \left[\eta \vert \nabla \rho \vert^2 + \frac{1}{\eta}\rho^2 \left(1-\rho\right)^2 \right]
\end{aligned}
\end{equation}
with $w$ a second weighting parameter that determines the effective penalty of maintaining an interface. To get sensible results, we have fixed $w=10^{-4}E_2$ with $E_2$ the Youngs modulus of the solid phase (see below). Because of the relative smallness and simplicity of the optimization problem, we have refrained from a hyperparameter optimization (e.g. \cite{Lynch2019}) but simply used educated guesses.

We use two different discretizations: A square unit cell discretized with a regular square grid or a hexagonal unit cell discretized with a regular hexagonal grid. In both cases, we use 31x31 grid points with two linear elements per pixel and linear shape functions. The discretized gradients are derived in detail in \cite{Leute2022} for the square grid and in~\ref{sec:gradient_hexagonal_grid} for the hexagonal grid. The material density is defined per grid point, so that both elements at one grid point always have the same material density.

The system consists of a hypothetical elastic, isotropic material (Youngs modulus $E_{2}$ and Poissons ratio $\nu_{2}=0$) and of void (Youngs modulus $E_1=0$ and Poissons ratio $\nu_1=0$) with a linear interpolation of the material properties in the interface, e.g. $E(\rho) = (E_2 - E_1) \rho + E_1$. We want to emphasize that the proposed equilibrium solver can handle elements with zero stiffness, so that we do not need to represent the void by a very weak material. Both materials are modelled with the standard 2-dimensional Hooke's law. The partial derivatives $\partial \t{\sigma} / \partial \t{\epsilon}$ and $\partial \t{\sigma} / \partial \rho$ can easily be calculated analytically.

We solve the equilibrium equation with the Newton-CG solver of \textsc{\textmu Spectre}. We optimize the aim function with a standard L-BFGS-B optimizer~\cite{Nocedal2006} as implemented in \textsc{scipy} with 0 and 1 as bounds on the material density. 

The initial phase distribution is a superposition of one-phase sine wave in the two dimensions or a random phase distribution, see Figure \ref{fig:top_opt_cubic} a and d respectively Figure \ref{fig:top_opt_hex} a and d.

\subsection{Validation of the sensitivity analysis}
\begin{figure}
    \includegraphics[width=0.95\textwidth]{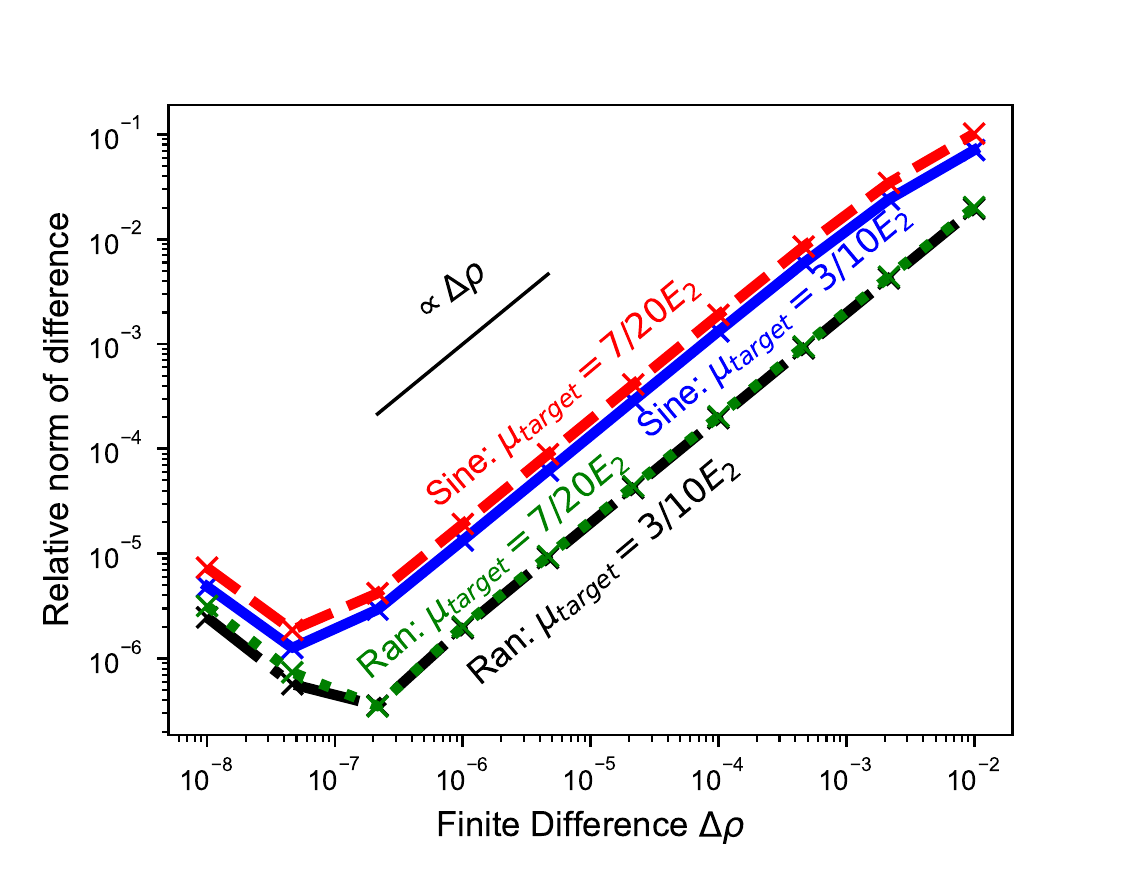}
    \caption{Difference between the sensitivity calculated with the adjoint method and the sensitivity calculated with the finite difference for a square grid discretization. The difference is shown for two different initial phases (random and sine wave) and two different target shear moduli $\mu_{\text{target}}$.}
    \label{fig:validation_adjoint_method}
\end{figure}

First, we validated the implemented adjoint method by a comparison with a finite difference calculation of the sensitivity. Figure \ref{fig:validation_adjoint_method} shows the norm of the difference between the finite difference calculation of the sensitivity and the sensitivity calculated with the discrete adjoint method for a square grid discretization. The difference is given for two different initial phases and two different target shear moduli. In each case, the norm decreases linearly with the finite difference $\Delta \rho$, until very small values of $\Delta \rho$ are reached. The results for a hexagonal grid (not shown) look similar.
We conclude that the finite difference calculation converges linearly towards the adjoint calculation until numerical errors get dominant, confirming our expressions and their implementation.

\subsection{Optimization results}
\begin{figure}
    \includegraphics[width=0.75\textwidth]{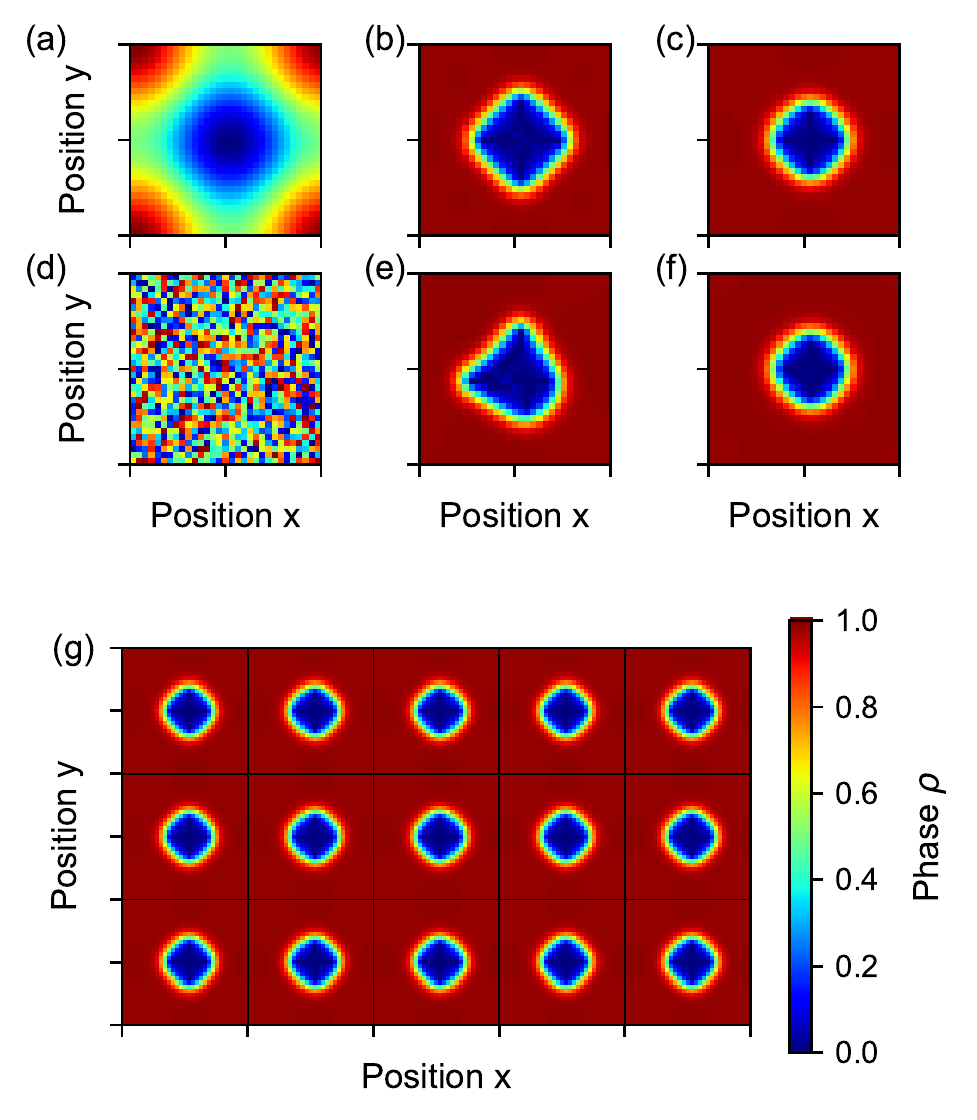}
    \caption{Result of the topology optimization for a square grid. The first row (a)-(c) shows the optimization for a sine wave as initial phase and the second row (d)-(f) the optimization for a random initial phase. The first column, (a) and (d), represents the initial phase distribution, the second column, (b) and (e) the optimized phase distribution for a target shear modulus of 3/10$E_2$ and the third column (c) and (f) the optimized phase distribution for a target shear modulus of 7/20 $E_2$. (g) shows several unit cells with the phase distribution of (c).}
    \label{fig:top_opt_cubic}
\end{figure}
\begin{figure}
    \includegraphics[width=0.9\textwidth]{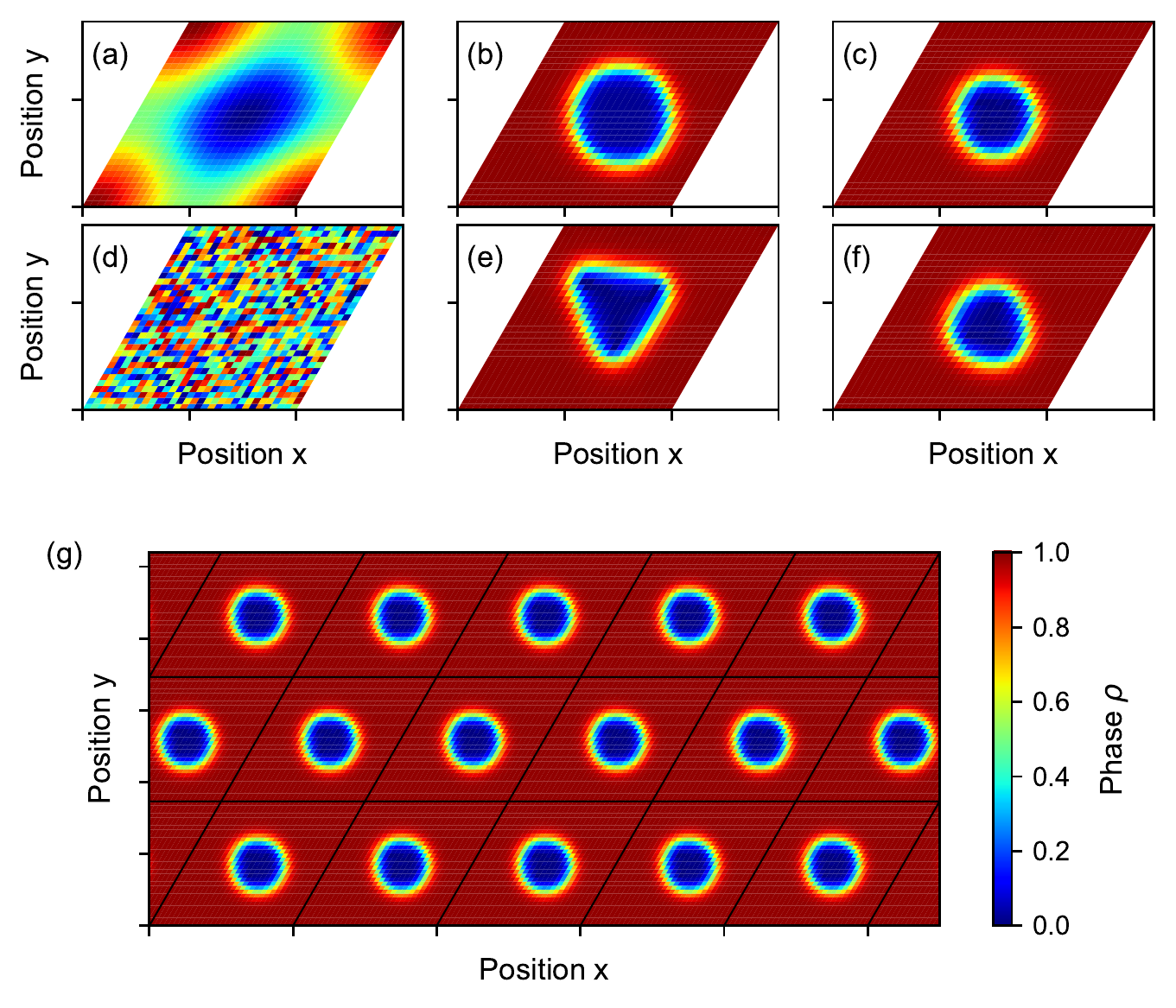}
    \caption{Result of the topology optimization for a hexagonal grid. The first row (a)-(c) shows the optimization for a sine wave as initial phase and the second row (d)-(f) the optimization for a random initial phase. The first column, (a) and (d), represents the initial phase distribution, the second column, (b) and (e) the optimized phase distribution for a target shear modulus of 3/10$E_2$ and the third column (c) and (f) the optimized phase distribution for a target shear modulus of 7/20$E_2$. (g) shows several unit cells with the phase distribution of (c).}
    \label{fig:top_opt_hex}
\end{figure}

The results of the optimizations with the square grid are shown in Fig.~\ref{fig:top_opt_cubic}, the results with the hexagonal grid in Fig.~\ref{fig:top_opt_hex}. Note that for comparability we have shifted the results of the different optimizations so that the hole is always in the center of the unit cell. We observe that the eight optimizations all lead to a single hole in the unit cell. For the sinus initial phase the hole is roughly square shaped in the case of the square grid (Fig. \ref{fig:top_opt_cubic} b and c) and circular in the case of the hexagonal grid (Fig. \ref{fig:top_opt_hex} b and c). For both grids, the hole is smaller for a larger target shear modulus. For the random initial phase and a target shear modulus of 7/20$E_2$, we get the same results as for the sinus initial phase for both grids (Fig.~\ref{fig:top_opt_cubic} f and Fig.~\ref{fig:top_opt_hex} f). However, for a random initial phase and target shear modulus of 3/10$E_2$ the hole is a quadrangle (Fig. \ref{fig:top_opt_cubic} e) in the case of the square grid and triangular (Fig. \ref{fig:top_opt_hex} e) in the case of the hexagonal grid.

The results of the optimization for a sine wave as initial phase agree with our expectations: Larger target shear modulus result in smaller holes and circular holes arranged in a triagonal hexagonal grid are isotropic with respect to the elasticity tensor \cite{Fil1964, Hu2000}. Since the target homogenized elastic constants were isotropic, the rectangular form of the holes in the case of the square grid compensates the anisotropy introduced by the square simulation cell. 

For $\mu_{\text{target}}=7/20E_2$, we get the same results for a random initial phase as for the sinus initial phase. For $\mu_{\text{target}}=3/10E_2$, the optimization finds other minimums with a random initial phase. Those are probably local minima.

\subsection{Auxetic metamaterials}
As a second example we want to optimize for an auxetic structure, i.e. a structure with a negative Poisson's ratio. The approach is the same as in section~\ref{sec:test_problem} with two changes:
Firstly, we prescribe a Poisson's ratio of $\nu_\text{target}=-1/3$ in addition to the target shear modulus $\mu_\text{target}=1/4E_2$. Secondly, we optimize only for the load case $\t{\epsilon}_1$ from section~\ref{sec:test_problem}.
With educated guessing we fixed the weighting parameters as $w=3\cdot 10^{-5}E_2$, all other parameters are the same as in section~\ref{sec:test_problem}.

\begin{figure}
    \includegraphics[width=0.9\textwidth]{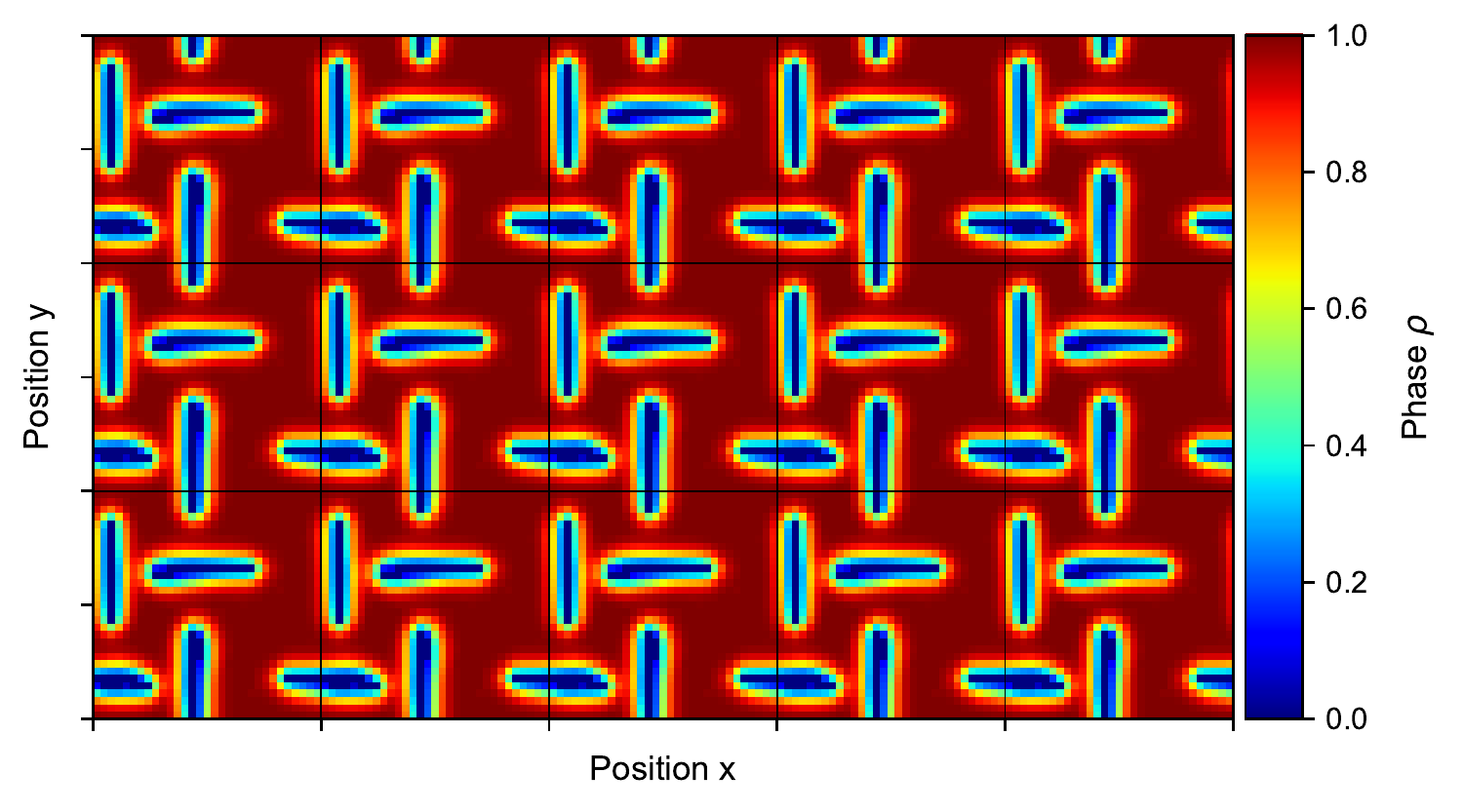}
    \caption{Result of the topology optimization for a target shear modulus of 1/4$E_2$ and a target Poisson's ration of -1/3, using a square grid discretization and starting from a random initial phase.}
    \label{fig:top_opt_negativ_poisson}
\end{figure}

Fig.~\ref{fig:top_opt_negativ_poisson} represents the result of the topology optimization for a random initial phase and a square grid discretization. It consists of a pattern of alternating horizontal and vertical slits. A similar pattern has been proposed by Taylor et al. to achieve auxetic structures with low porosity~\cite{Taylor2014}. We conclude that the proposed method can be used for the design of simple auxetic structures.

\section{Conclusion}


In this paper, we have developed the theory for the discrete adjoint method in the context of Fourier-based micromechanics with compatibility projection. We implemented and validated this method within \textsc{\textmu Spectre}~\cite{muspectre}. Our optimization examples with inclusions of zero stiffness demonstrate that this method can be used for the design of simple metamaterials. Furthermore, the concept of compatibility projection can be extended to construct preconditioners for standard finite element formulations. Our derived expressions are extremely simple and efficient to calculate within existing solution frameworks, fostering the way for the rational design of complex composites or metamaterials.

\section*{Acknowledgements}

We thank W. Beck Andrews, Andrea Codrignani, Martin Ladecký, Ivana Pultarová, Antoine Sanner and Jan Zeman for enlightening discussion. We acknowledge funding by the Carl Zeiss Foundation (Research cluster ``Interactive and Programmable Materials - IPROM''), the European Research Council (StG-757343), the Deutsche Forschungsgemeinschaft (EXC 2193/1 - 390951807) and the Swiss National Science Foundation (Ambizione grant 174105).

\appendix
\section{Discrete gradient for a hexagonal grid}
\label{sec:gradient_hexagonal_grid}
\begin{figure}
    \includegraphics[width=0.7\textwidth]{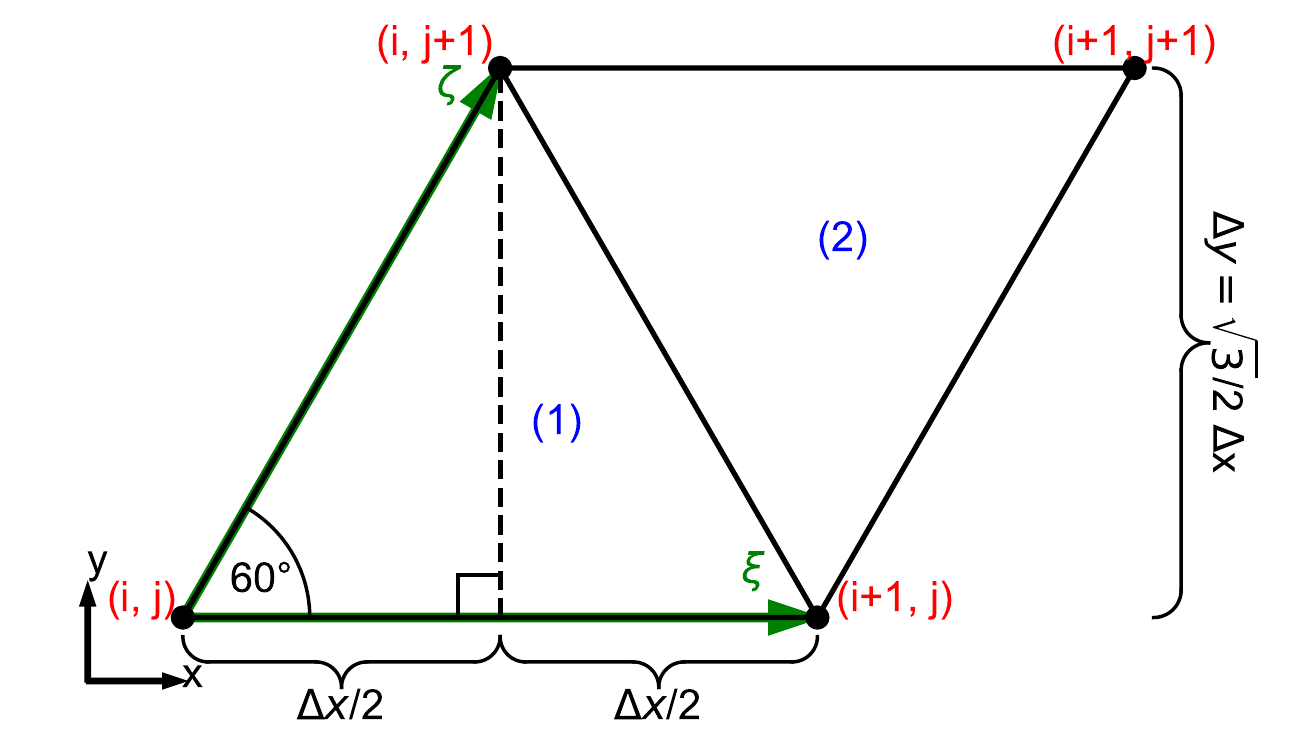}
    \caption{Hexagonal unit cell}
    \label{fig:hex_unit_cell}
\end{figure}

In this appendix we derive the gradient for a hexagonal grid and linear shape functions. The unit cell of this grid is represented in Fig.~\ref{fig:hex_unit_cell}.
The Cartesian coordinate system $\left(x, y \right)$ and the element system $\left(\xi, \zeta \right)$ are connected by:
\begin{align*}
    & \xi = \frac{x}{\Delta x} - \frac{y}{2\Delta y}\\
    & \zeta = \frac{y}{\Delta y}
\end{align*}
We use the linear shape functions:
\begin{align*}
    & N_{00}^{(1)}\left( \xi, \zeta\right) = 1 - \xi - \zeta && N_{10}^{(2)}\left( \xi, \zeta\right) = 1 - \zeta \\ 
    & N_{10}^{(1)}\left( \xi, \zeta\right) = \xi &&
    N_{01}^{(2)}\left( \xi, \zeta\right) = 1 - \xi \\
    & N_{11}^{(1)}\left( \xi, \zeta\right) = \zeta &
    & N_{11}^{(2)}\left( \xi, \zeta\right) = \xi + \zeta - 1\\
\end{align*}
where the superscript designates the number of the triangular element (see Fig.~\ref{fig:hex_unit_cell}). 
The partial derivatives of a function $f$ follow directly as:
\begin{align*}
    \frac{\partial f^{(1)}}{\partial x} &= \frac{f_{i+1,j}^{(1)} - f_{i,j}^{(1)}}{\Delta x}
    \\
    \frac{\partial f^{(2)}}{\partial x} &= \frac{f_{i+1,j+1}^{(2)} - f_{i,j+1}^{(2)}}{\Delta x}\\
    \frac{\partial f^{(1)}}{\partial y} &= \frac{2f_{i,j+1}^{(1)} - f_{i,j}^{(1)} - f_{i+1,j}^{(1)}}{2 \Delta y}
    \\
    \frac{\partial f^{(2)}}{\partial y} &= \frac{-2f_{i+1,j}^{(2)} + f_{i,j+1}^{(2)} + f_{i+1,j+1}^{(2)}}{2 \Delta y}
\end{align*}
with $f_{i,j}^{(a)}$ the value of $f$ at the grid point $i$, $j$ and the triangular element $a$.

%

\end{document}